\documentclass{ws-procs10x7}
\usepackage{euscript}
\usepackage{amsmath}
\usepackage{amssymb}
\usepackage{cite}
\usepackage{color}

\def\mathswitchr#1{\relax\ifmmode{\mathrm{#1}}\else$\mathrm{#1}$\fi}

\newcommand {\pslash}{\hbox{$\not\hbox{\kern-2.3pt $p$}$}}

\def\rQCED{{\rm QCED}}

\def\alf1{ {\alpha\over\pi} }
\def\rQCD{{\rm QCD}}
\begin{document}

\title{Threshold Corrections in QED$\otimes$QCD at the LHC}

\author{B.F.L. Ward }

\address{Department of Physics, Baylor University, Waco, TX, USA}

\author{C. Glosser}

\address{Department of Physics, Southern Illinois University, Edwardsville, IL,  USA}  
\author{S. Jadach}

\address{Institute of Nuclear Physics, Krak\'ow, Poland\\
Theory Division, CERN, Geneva, Switzerland} 

\author{S. A. Yost}

\address{Department of Physics, Baylor University, Waco, TX, USA} 

\twocolumn[\maketitle\abstract{In some processes at the LHC,
theoretical precisions of 1\% are desired. With an eye toward
such precisions, we introduce the 
theory of the simultaneous YFS resummation of QED and QCD
to compute the size of the
expected resummed soft radiative threshold effects in 
precision studies of heavy
particle production at the LHC. Our results, 
that the soft QED threshold effects are
at the level of 0.3\% whereas the soft QCD threshold effects enter at the 
level of 20\%, show that both must be controlled to be on 
the conservative side to achieve such precision goals.}]

\section{Introduction}
In high energy collider processes, such as t\=t production at
FNAL, precision predictions for soft multiple gluon (n(g)) effects
are already needed: the uncertainty on $m_t$~\cite{mterror}, 
$\delta m_t = 4.3$ GeV, receives a soft n(g) uncertainty $\sim$ 2-3 GeV,
for example. At the LHC/ILC,
the requirements will be even more demanding
and soft n(g) MC exponentiation results will be an 
important part of the necessary theory -- YFS exponentiated
${\cal O}(\alpha_s^2)L$ calculations, in the presence of parton showers, 
on an event-by-event basis.

How relevant are QED higher order corrections when QCD is
controlled at $\sim 1\%$ precision? Many authors~\cite{qcdlit} 
are preparing the
necessary results that would lead to such a precision on QCD
for LHC processes.
Estimates by Refs.~\cite{cern2000,spies,james1,roth,james2}
show that one gets few per mille
effects from QED corrections to structure function evolution.
The well-known possible enhancement of QED corrections at threshold,
especially in resonance production, leads us to estimate
how big are these effects at the LHC.

We treat QED and QCD simultaneously in the respective 
YFS~\cite{yfs,yfs1} exponentiation
to estimate the role of the QED threshold effects at the LHC 
in the representative processes
$pp\rightarrow V +n(\gamma)+m(g)+X\rightarrow \bar{\ell} \ell'
+n'(\gamma)+m(g)+X$, where 
$V=W^\pm,Z$,and $\ell = e,\mu,~\ell'=\nu_e,\nu_\mu ( e,\mu )$
respectively for $V=W^+ ( Z )$, and  
$\ell = \nu_e,\nu_\mu,~\ell'= e,\mu$ respectively for $V = W^-$.
Precision studies of these processes have been proposed for
luminometry at the LHC~\cite{lhclum} and at FNAL~\cite{fnallum}, 
where 2-3\% is the target
precision tag for the LHC, for example. The latter would indeed require
a theoretical precision tag of $\sim 1\%$ in order that
the theory error not figure too prominently in the over-all 
precision. 

Our discussion is organized as follows. After giving a brief
review of the YFS theory and its extension to QCD in the next section,
in Section 3 we introduce QED$\otimes$QCD YFS exponentiation.
In Section 4, we apply the new development to the 
threshold corrections in single
V production at the LHC and at FNAL. Section 5 contains some
concluding remarks.\par

\section{Review of the YFS Theory and its Extension to QCD}

As realized in Refs.~\cite{yfs1} by Monte Carlo methods,
for $e^+(p_1)e^-(q_1)\rightarrow \bar{f}(p_2) f(q_2) +n(\gamma)(k_1,\cdot,k_n)$,
renormalization group improved YFS theory~\cite{bflw1yfs} gives,{\small
\begin{eqnarray}
d\sigma_{exp}&=e^{2\alpha\,Re\,B+2\alpha\,
\tilde B}\sum_{n=0}^\infty{1\over n!}\int\prod_{j=1}^n{d^3k_j\over k_j^0
}\nonumber\\&\int {d^4y\over(2\pi)^4}e^{iy(p_1+q_1-p_2-q_2-\sum_jk_j)+D}\nonumber\\
&\qquad\bar\beta_n(k_1,\dots,k_n){d^3p_2d^3q_2\over p_2^0q_2^0}
\label{eqone1}\end{eqnarray}}\noindent
where the YFS infrared functions $\tilde B,~B$ and $D$ are
known. 
For example, the YFS hard photon residuals $\bar\beta_i$ in (\ref{eqone1}), 
$i=0,1,2$,
are given in the first paper in Ref.~\cite{yfs1} and realize the
YFS exponentiated exact ${\cal O}(\alpha)$
and LL ${\cal O}(\alpha^2)$
cross section for Bhabha scattering via a
corresponding Monte Carlo realization
of (\ref{eqone1}).

In Refs.~\cite{qcdref,dglapsyn}
we have extended the YFS theory to QCD:{\small
\begin{equation}
\begin{split}
d\hat\sigma_{\rm exp}&= \sum_n d\hat\sigma^n 
         =e^{\rm SUM_{IR}(QCD)}\sum_{n=0}^\infty\int\prod_{j=1}^n{d^3
k_j\over k_j}\\&\quad\int{d^4y\over(2\pi)^4}e^{iy\cdot(P_1+P_2-Q_1-Q_2-\sum k_j)+
D_\rQCD}\\
&*\tilde{\bar\beta}_n(k_1,\ldots,k_n){d^3P_2\over P_2^{\,0}}{d^3Q_2\over
Q_2^{\,0}}
\end{split}
\label{subp15a}
\end{equation}}\noindent
where gluon residuals 
$\tilde{\bar\beta}_n(k_1,\ldots,k_n)$
, defined by Ref.~\cite{qcdref}, 
are free of all infrared divergences to all 
orders in $\alpha_s(Q)$. The functions
$SUM_{IR}(QCD), D_\rQCD$, together with 
the basic infrared functions 
$B^{nls}_{QCD},{\tilde B}^{nls}_{QCD},{\tilde S}^{nls}_{QCD}$ 
are specified in Ref.~\cite{qcdref}.
We call attention to the essential
compensation between
the left over genuine non-Abelian IR virtual and real singularities
between $\int dPh\bar\beta_n$ and $\int dPh\bar\beta_{n+1}$ respectively
that really allows us to isolate $\tilde{\bar\beta}_j$ and distinguishes
QCD from QED, where no such compensation occurs.

We stress that the YFS resummation which we exhibit here is
fully consistent with that of Refs.~\cite{ster,cattrent}.
We refer the reader to Ref.~\cite{CG1} for more discussion of this point.

\section{Extension to QED$\otimes$QCD and QCED}

Simultaneous exponentiation of
QED and QCD higher order effects~\cite{CG1}
gives{\small
\begin{eqnarray}
B^{nls}_{QCD} \rightarrow B^{nls}_{QCD}+B^{nls}_{QED}\equiv B^{nls}_{QCED},\cr
{\tilde B}^{nls}_{QCD}\rightarrow {\tilde B}^{nls}_{QCD}+{\tilde B}^{nls}_{QED}\equiv {\tilde B}^{nls}_{QCED}, \cr
{\tilde S}^{nls}_{QCD}\rightarrow {\tilde S}^{nls}_{QCD}+{\tilde S}^{nls}_{QED}\equiv {\tilde S}^{nls}_{QCED}
\label{irsub}
\end{eqnarray}} 
which leads to {\small
\begin{equation}
\begin{split}
d\hat\sigma_{\rm exp} &= e^{\rm SUM_{IR}(QCED)}\\
   &\sum_{{n,m}=0}^\infty\int\prod_{j_1=1}^n\frac{d^3k_{j_1}}{k_{j_1}} 
\prod_{j_2=1}^m\frac{d^3{k'}_{j_2}}{{k'}_{j_2}}
\int\frac{d^4y}{(2\pi)^4}\\&e^{iy\cdot(p_1+q_1-p_2-q_2-\sum k_{j_1}-\sum {k'}_{j_2})+
D_\rQCED} \\
&\tilde{\bar\beta}_{n,m}(k_1,\ldots,k_n;k'_1,\ldots,k'_m)\frac{d^3p_2}{p_2^{\,0}}\frac{d^3q_2}{q_2^{\,0}},
\end{split}
\label{subp15b}
\end{equation}}\noindent
where  the new YFS residuals, defined in Ref.~\cite{CG1}, 
$\tilde{\bar\beta}_{n,m}(k_1,\ldots,k_n;k'_1,\ldots,k'_m)$, with $n$ hard gluons and $m$ hard photons,
represent the successive application
of the YFS expansion first for QCD and subsequently for QED. The
functions ${\rm SUM_{IR}(QCED)},D_\rQCED$ are determined
from their analoga ${\rm SUM_{IR}(QCD)},D_\rQCD$ via the
substitutions in (\ref{irsub}) everywhere in expressions for the
latter functions given in Refs.~\cite{qcdref}. 

Infrared Algebra(QCED): the average Bjorken $x$ values
\begin{eqnarray}
x_{avg}(QED)&\cong \gamma(QED)/(1+\gamma(QED))\nonumber\\
x_{avg}(QCD)&\cong \gamma(QCD)/(1+\gamma(QCD))
\nonumber
\end{eqnarray}
where
$\gamma(A)=\frac{2\alpha_{A}{\cal C}_A}{\pi}(L_s
-1)$, $A=QED,QCD$, with
${\cal C}_A=Q_f^2, C_F$, respectively, for 
$A=QED,QCD$ and the big log $L_s$, imply that
QCD dominant corrections happen an
order of magnitude earlier than those for QED.This means that
that the leading $\tilde{\bar\beta}_{0,0}^{(0,0)}$-level
gives a good estimate of the size of the effects we study.

\section{QED$\otimes$QCD Threshold Corrections at the LHC} 

We shall apply the new simultaneous QED$\otimes$QCD exponentiation
calculus to the single Z production with leptonic decay 
at the LHC ( and at FNAL)
to focus on the ISR alone, for definiteness.
See also the work of Refs.~\cite{baurall,ditt,russ} for exact ${\cal O}(\alpha)$ results and Refs.~\cite{van1,van2,anas}
for exact ${\cal O}(\alpha_s^2)$ results.

For the basic formula (we use the standard notation here~\cite{CG1}){\small
\begin{eqnarray}
d\sigma_{exp}(pp\rightarrow V+X\rightarrow \bar\ell \ell'+X') =\nonumber\\
\sum_{i,j}\int dx_idx_j F_i(x_i)F_j(x_j)d\hat\sigma_{exp}(x_ix_js),
\label{sigtot} 
\end{eqnarray}}\noindent
we use the result in (\ref{subp15b}) here with semi-analytical
methods and structure functions from Ref.~\cite{mrst1}.
A Monte Carlo realization will appear elsewhere~\cite{elsewh}.

We {\bf do not} attempt to replace HERWIG~\cite{herwig} 
and/or PYTHIA~\cite{pythia} --
we intend to combine our exact YFS calculus with HERWIG and/or PYTHIA
{\bf by using the latter in lieu of the $\{F_i\}$}.
This combination of theoretical constructs can be 
systematically improved with
exact results order-by-order in $\alpha_s$, where  
currently the state of the art in such
a calculation is the work of Frixione and Webber in Ref.~\cite{frixw}
which accomplishes the combination of an exact ${\cal O}(\alpha_s)$
correction with HERWIG. We note that, even in this latter result,
the gluon azimuthal angle is averaged in the combination. We note that
the recent alternative parton shower algorithm by Jadach and 
Skrzypek in Ref.~\cite{jadskrz} can also be used in our theoretical
construction here. Due to its lack of the appropriate color coherence~\cite{mmm}, we do not consider ISAJET~\cite{isajet} here.

We compute , with and without QED, the ratio
$r_{exp}=\sigma_{exp}/\sigma_{Born}$
to get the results
(We stress that we {\em do not} use the narrow resonance approximation here.){\small
\begin{equation}
r_{exp}=
\begin{cases}
1.1901&, \text{QCED}\equiv \text{QCD+QED,~~LHC}\\
1.1872&, \text{QCD,~~LHC}\\
1.1911&, \text{QCED}\equiv \text{QCD+QED,~~Tevatron}\\
1.1879&, \text{QCD,~~Tevatron.}\\
\end{cases}
\label{res1}
\end{equation}}
We see that QED is at the level of .3\% at both LHC and FNAL.
This is stable under scale variations~\cite{CG1}.
We agree with the results in Refs.~\cite{baurall,ditt,russ,van1,van2}
on both of the respective sizes of the QED and QCD effects.
The QED effect is similar in size to structure function
results in Refs.~\cite{cern2000,spies,james1,roth,james2}.

\section{Conclusions}

YFS theory (EEX and CEEX) extends to 
non-Abelian gauge theory and allows simultaneous exponentiation of 
QED and QCD. For QED$\otimes$QCD we find that
full MC event generator realization is possible in a way that
combines our calculus with Herwig and Pythia in principle.
Semi-analytical results for QED (and QCD) threshold effects agree 
with literature on Z production. As QED is at the .3\% level, 
it is needed for 1\% LHC theory predictions.
A firm basis for the complete ${\cal O}(\alpha_s^2,\alpha\alpha_s,\alpha^2)$
results needed for the FNAL/LHC/RHIC/TESLA/ILC physics 
has been demonstrated and all of the latter are in progress.

\section*{Acknowledgments}
Two of us ( S.J. and B.F.L.W.)
thank Profs. S. Bethke and L. Stodolsky for the support and kind
hospitality of the MPI, Munich, while a part of this work was
completed. This work was supported partly by US DoE contract
DE-FG05-91ER40627 and by NATO grants PST.CLG.97751,980342.

\end{document}